\documentclass[%
 reprint,
 amsmath,amssymb,
 aps,
]{revtex4-1}

\usepackage{graphicx}
\usepackage{dcolumn}
\usepackage{bm}

\begin{document}

\preprint{APS/123-QED}

\title{Neutrino interactions at COMET}

\author{D.~Mulmule}
\affiliation{%
 Department of Physics, Indian Institute of Technology Bombay, Main Gate Rd, IIT Area, Powai, Mumbai, Maharashtra 400076 \\
}%


\date{\today}

\begin{abstract}
  The experimental search for coherent neutrinoless conversion of muon to electron in the presence of a nucleus, aims to probe the possibility of charged lepton flavour violation. The COMET experiment at J-PARC is one such setup offering unprecedented statistical reach, with single event sensitivities down to $\mathcal{O}$ ($\mathrm{10^{-17}}$). Its successor experiment - PRISM, planned as the final stage, is expected to take this bound further down by two orders of magnitude. The electrons from standard model decay of bound muons are considered the most formidable physics background to the detection of these conversion electrons. Keeping in view the high stopped muon statistics for COMET and PRISM, the rate of charged current interaction of decay neutrinos with nuclear protons is also non-negligible. In this work, we perform a calculation of the positron event rates expected due to interactions of electron antineutrinos with Al target protons in the COMET setup. About $\mathrm{7\pm1}$ such $\mathrm{e^{+}}$ events per $\mathrm{10^{18}}$stopped muons are expected.

\end{abstract}

\maketitle


\section{Introduction}
The discovery of neutrino oscillations implied two significant departures from the standard model (SM) of particle physics, first being the existence of neutrino masses, and the second, the violation of individual lepton flavours. To explain the oscillations, SM had to be minimally extended to include heavy Dirac right-handed neutrinos as counterparts to extremely light ($\lesssim$ $\mathcal{O}$(eV)) left handed neutrinos~\cite{1ASAKA}. This extension entailed the possibility of rare `charged lepton flavour violation' (CLFV) transitions at loop level, but with extremely tiny branching ratio (B.R.), for e.g. for $\mathrm{{\mu}^{+} \rightarrow e^{+}\gamma}$ the B.R. in this minimally extended model is $\mathcal{O}$ $\mathrm{(10^{-54})}$ ~\cite{2petcov}. However, CLFV processes are postulated with an enhanced B.R. by some BSM theories such as supersymmetry, large extra dimensions and extended Higgs sectors etc.~\cite{3Calibbi,4cirigliano}. The CLFV B.R. postulated by such theories is within reach of the current generation accelerator based experiments. The $\mathrm{{\mu}^{-}N \rightarrow e^{-}N^{'}}$ conversion process is the most ideally suited for experiments due to its fixed energy conversion-electron signal at the muon energy endpoint. The COMET experiment at J-PARC~\cite{5Abramishvilli}
is one of the fore-runners of this search, with its higher statistical reach allowing for rare event searches other than CLFV conversion~\cite{7Mihara}. Al metal has been chosen as the target element for stopping muons, with possible use of Ti in future. Al offers the best combination of endpoint energy and decay time suitable for the experimental setup. The projected single event sensitivity (SES) is as low as $\mathrm{\sim 2 \times 10^{-17}}$ thus making many of the BSM predictions testable. The signals due to the 100 MeV and above electron-like events from non-conversion physics form the potential background to these BSM physics processes. The major backgrounds are classified based on their sources, with the most formidable being the intrinsic physics background due to known SM processes associated with stopped muons. The other two sources are either beam-related or those due to cosmic-rays and misidentification of tracks. Among the intrinsic backgrounds are 1) Decay-In-Orbit (DIO) electrons 2) Radiative Muon Capture (RMC) $\gamma$ s and 3) proton/neutron emission from nucleus following capture of bound muon. The DIO case is particularly interesting as both the bound muon and decay electron can interact electromagnetically with the nucleus and stretch the decay endpoint energy of electron very close to the muon rest mass. The neutrinos from the DIO event are not usually accounted in these considerations, but, in experiments designed for higher statistics of stopped muons, there is indeed a possibility of observing a charged current interaction of these decay neutrinos with target protons. In the following sections this possibility is explored with special emphasis on the particulars of the COMET experiment. In section II we discuss the muon decay and the decay neutrino characteristics and interactions. Section III discusses the rate calculations for their interactions with Al nucleus.

\section{Muon decay neutrinos and their interactions}

In SM, the muon undergoes flavour conserving three body decay, as follows:
\begin{equation}
{\mu}^{-} \rightarrow e^{-} + \nu_{\mu} + \overline{\nu_{e}}
\end{equation}
The three-body kinematics dictates that, for a decay-at-rest (DAR) muon the daughter particles share the available rest mass energy of muon ($\mathrm{M_{\mu}}$) and can have upto a maximum of 52.8 MeV ($\mathrm{\frac{M_{\mu}}{2}}$). Since, the experiments are concerned with the possibility of CLFV conversion, only $\mathrm{{\mu}^{-}}$ beam is used to allow a positively charged target nuclei to bind it in its orbit. Once the muon gets bound by the nuclear field it cascades down to a tighter 1s orbit with a radius $\mathrm{\sim \frac{1}{207}}$ times of that for $\mathrm{{e}^{-}}$, as it is about 207 times more massive. This transition happens within $\mathrm{\sim 10^{-13}}$ seconds. There are two competing SM processes for the muon to undergo in this state 1) Decay-in-orbit (DIO), which happens about 40$\%$ of the times and 2) Nuclear capture the remaining 60$\%$. This reduces the lifetime of muon from 2.2 $\mathrm{\mu}s$ to less than 1 $\mathrm{\mu}s$, and in the particular case of Al, it is 864 ns. The proximity to the nucleus means that
the DIO electrons can undergo EM interactions and have higher energies upto the total muon mass minus the few keV binding energy (B.E.) of the 1s orbit. Meanwhile, neutrinos from the decay don't feel any such force and are free streaming particles once produced. There energies still follow the original distribution upto 52.8 MeV~\cite{9EITEL} and average energy is $\sim$30 MeV; see solid blue histogram in figure~\ref{fig:Overlay}. In this energy range the muon neutrino can't produce a muon but an electron anti-neutrino has sufficient energy to produce positrons through inverse beta decay (IBD) interaction with a proton as follows:
\begin{equation}
\overline{\nu_{e}} + {p}^{+} \rightarrow e^{+} + n^{0}
\end{equation}
The positron then essentially takes away all the kinetic energy of the $\overline{\nu_{e}}$ for energies above the threshold for IBD i.e. 1.806 MeV. These positrons can potentially form a background on their own or through annihilation photons traveling further from the stopping target. On the other hand, such emissions can be studied as probes of $\overline{\nu_{e}}$ induced transition between nuclear states. The newly produced $\mathrm{{}^{27}Mg}$ is a beta decaying isotope with half-life of a few minutes, although, the endpoint energies and de-excitation $\gamma$ energies are only limited upto 2 MeV. These reactions can be of interest for nuclear, astrophysical and weak interaction studies.

\section{Calculation of positron event rate at COMET }
The calculations for neutrino interactions presented here apply to the operation of COMET for sensitivities comparable to or better than $\mathrm{10^{-17}}$ which is possible in its phase-II operation. Also these calculations can be scaled for the successor - `Phase Rotated Intense Slow Muons' (PRISM) experiment with projected SES $\mathrm{\sim 10^{-19}}$. For a full run of Phase-II operation of COMET an estimated $\mathrm{\sim 1 \times 10^{18}}$ number of muons are expected to be stopped in the target.
\begin{figure}[h]
  \includegraphics[width=\linewidth]{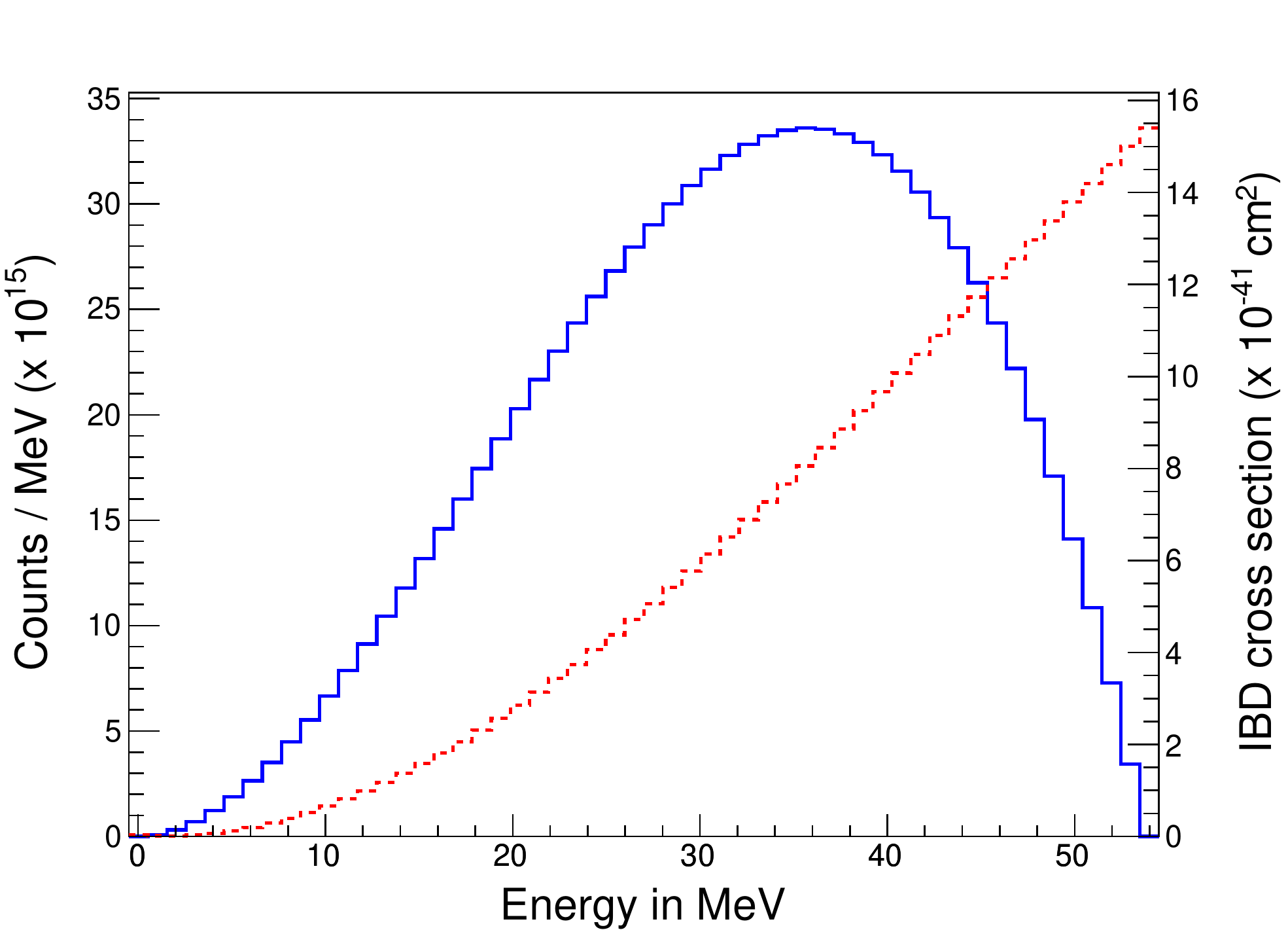}
  \caption{ Spectrum of $\mathrm{\mu}$-DAR neutrinos in blue (solid) histogram corresponding to 1 year of phase-II run statistics (left Y-axis scale) overlayed with IBD cross section in red (dashed) histogram for $\overline{\nu_{e}}$ (right Y-axis scale).}
  \label{fig:Overlay}
\end{figure}
As seen earlier, $\sim$40$\%$ are DIO candidates, reducing the potential $\overline{\nu_{e}}$ sources to $\mathrm{\sim 0.4 \times 10^{18}}$. 
The $\overline\nu_{e}$ spectra and IBD cross section overlayed upon each other are shown in the figure~\ref{fig:Overlay}.
The orbit radius for bound muon can be approximated using the reduced mass coefficient multiplied to the Bohr radius of a 1s electron. As per the formula, for muons with mass `$\mathrm{m_{\mu}}$' and charge `e' the modified Bohr radius formula becomes:\\
\begin{equation*}
  r_{\mu} = (m_{e} / m_{R}) \times r_{e} \mathrm{~\r A}
\end{equation*}
where $\mathrm{m_{R} = m_{\mu} \times m_{Nucleus} / (m_{\mu} + m_{Nucleus})}$ and $\mathrm{r_{e} = a_{0} \times n^{2} / Z ~\r A}$ ( $\mathrm{a_{0} = 0.529 ~\r A}$: Bohr's radius for electron in H atom). Substituting these we get the radius of 1s orbit of muon in Al (Z=13) as: \\
\begin{equation}
  r_{\mu} = 0.0048 \times 0.529 \times (1/13) \mathrm{\r A \sim 19.6~fm}
\end{equation}

The inverse beta decay cross section for electron anti-neutrinos for folding with the $\mathrm{\mu}$-DAR neutrino spectrum is taken from reference~\cite{11Vogel}. The number of interactions (events) for one year of phase-II operation is given by:
\begin{equation*}
  W = N_{target} \times \varphi(E_{\nu}) \times (\frac{{\sigma_{IBD}}({E_{\nu}})}{A}) 
\end{equation*}\\
where, $\mathrm{N_{target}}$ is the number of target protons (13 for Al) available for each IBD interaction, 
$\mathrm{\varphi(E_{\nu})}$ is the decay neutrino fluence over one year of operation ($\mathrm{\sim 0.4 \times 10^{18}}$) and the factor $\mathrm{\frac{{\sigma}_{IBD}(E_{\nu})}{A}~is~actually~a~integral~\iiint \frac{{\sigma}_{IBD}(E_{\nu})}{4 \times \pi \times D(r)^{2}}}$ in spherical co-ordinates, representing the probability of occurence of IBD with an Al proton, $\mathrm{\sigma_{IBD}}$ being the IBD cross-section and $\mathrm{D(r)^{2}}$ the stand-off distance. We neglect the possibility of neutrinos from different muonic atom decays impacting one another as the distance scale increases from fermi to angstrom levels and hence by inverse square law the intensity drops drastically as we move further away from one interaction center to another. For a single  $\overline{\nu_{e}}$-nucleus pair, we consider a non-negligible gaussian spatial distribution of protons and the $\mathrm{\langle D(r)^{2} \rangle}$ comes out to be $\mathrm{\sim r_{\mu}^{2}}$ with deviation only in the 3rd decimal place. The radius of the Al nucleus needed for the above calculation is $\sim$ 3.6 fm (using $\mathrm{r_{nucleus} = 1.2 \times (A)^{1/3} fm}$). The quantities $\mathrm{\varphi}$ and $\mathrm{\sigma_{IBD}}$ are both functions of neutrino energy and hence folded to obtain output spectrum of positrons.
\begin{figure}[h]
  \includegraphics[width=\linewidth]{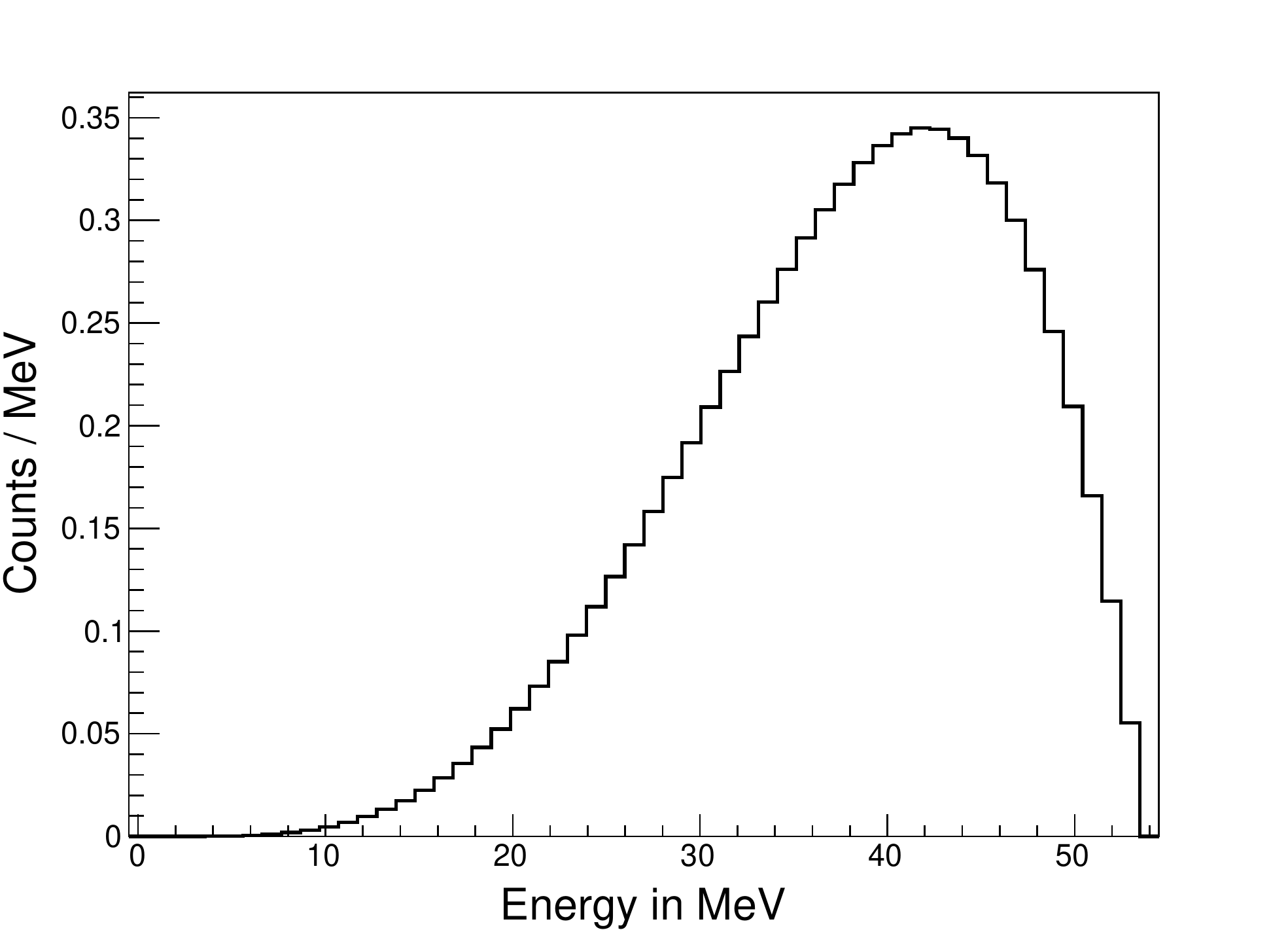}
  \caption{ Spectrum of the outgoing positrons.}
  \label{fig:final}
\end{figure}
The shape of the spectrum, as seen in figure~\ref{fig:final}, is obtained assuming isotropic neutrino flux. This assumption is based upon the argument that $\mathrm{d\Gamma/dcos\theta}$ = 1 when $P_{\mu}$ - the polarisation of muon is $\sim$0. This is true for orbit muons, as the drop of muon to the lowest orbit completely depolarises it. The final IBD rate from the above calculation comes out to be $\mathrm{7\pm1~e^{+}~events~per~10^{18}}$ stopped muons, assuming a total uncertainty of upto 15$\%$ due to uncertainties in stopped muon numbers, cross section and distances involved. A scaling up of the stopped muon numbers as anticipated for future experiments, will scale this rate proportionally. Before concluding the discussion it is necessary to point out that one of the key components to be added in the phase-II operation of COMET is the electron spectrometer for selecting only the high momentum $e^{-}$s ($>$70 MeV/c) rejecting most of the DIO electrons through use of a dipole magnet and DIO blockers. But for a relatively high energy positrons (or $\gamma$s) the dipole field and blocker behavior needs to be studied. Further, if the beam intensity is ramped up to phase-II levels while operating the phase-I setup there is higher chance of observing such events due to the absence of selective magnetic field. Thus, scanning for IBD positrons reaching the cylindrical detector setup (as designed for phase-I operation) can be a worthwhile exercise.
\section{Conclusion}
The COMET experiment at J-PARC to probe the CLFV is expected to start full Phase - I operation in 2022 followed by Phase-II start about a year later. The final target is to reach CLFV sensitivities 4 orders below the current limit through probe of nearly $\mathrm{\sim 1 \times 10^{18}}$ stopped muon events/year. The decay of these bound muons leads to a projected estimate of $\mathrm{7\pm1}$ IBD induced $e^{+}$ events/yr coming from interactions with protons in Al nucleus. This $<$100 MeV IBD interaction data may be useful simply for a background component estimate or can possibly be used for a low energy neutrino interaction study in context of nuclear-astrophysics.\\

 The author is thankful to Prof. S. Umasakar(IIT-B) for enlightning discussions on this topic.

\bibliography{COMET_PRLFormat}

\end{document}